\documentstyle[12pt]{article}
\newcommand{\apm}{\!\stackrel{\leftrightarrow\;}{\partial_{\mu}}\!}
\newcommand{\apn}{\!\stackrel{\leftrightarrow\;}{\partial'_{\nu}}\!}
\newcommand{\cint}{\int_{\Sigma}d\Sigma}
\title{A general-covariant concept of particles in curved background}
\author{Hrvoje Nikoli\'c \\
Theoretical Physics Division, Rudjer Bo\v{s}kovi\'{c} Institute, \\
P.O.B. 180, HR-10002 Zagreb, Croatia \\
{\normalsize hrvoje@faust.irb.hr} \\
\makebox[1in]{} \\
}
\date{\today}
\begin{document}
\maketitle
\begin{abstract}
A local current of particle density for scalar 
fields in curved background is constructed. 
The current depends on the choice of a two-point 
function. There is a choice that leads to local non-conservation 
of the current in a time-dependent gravitational background, 
which describes local particle production consistent with the usual 
global description based on the Bogoliubov transformation. 
Another choice, which might be the most natural one, leads to 
the local conservation of the current.
\end{abstract}
\vspace*{0.5cm} 
PACS: 04.62.+v; 11.10.-z \newline
Keywords: Current of Particle Density; Curved Background; Particle 
Production
\vspace*{0.9cm}

\noindent
One of the main problems regarding quantum field theory in curved space-time is 
how to introduce the concept of particles. 
The problem is related to the fact 
that particles are {\em global} objects in the conventional
approach \cite{ful,bd}, while covariance with respect to general coordinate
transformations requires local objects.
This led some experts to 
believe that only local operators in field theory  
were really meaningful and, consequently, that the concept of particles  
did not have any fundamental meaning \cite{davies}.
However, particles are what we observe in experiments. Moreover, from 
the experimental point of view, nothing is more local than the 
concept of particles. If we require that quantum field theory 
describes the observed objects, then it should  
describe particles as local objects. In this letter 
we show that the concept of particles can be introduced in a 
local and covariant manner. 

Our approach is based on a similarity 
between the number of particles and charge. For complex fields, 
the total number of particles is the sum of the number of particles and 
antiparticles, while the total charge is the difference of these two 
numbers. The concept of charge can be described in a local 
and covariant manner because there exists a local vector current 
of charge density. We find that a similar vector current exists 
for the number of particles as well. Nevertheless, it appears that this 
local current is not unique, but depends on the choice of a   
two-point function $W(x,x')$. When a unique 
(or a preferred) vacuum $|0 \rangle$ exists, 
then $W(x,x')$ is equal to the Wightman function  
$\langle 0|\phi(x)\phi(x')|0 \rangle$ and the current is conserved.
When such a vacuum does not exist, then there is a
choice of $W$ that leads to local non-conservation
of the current in a time-dependent gravitational background,
which describes local particle production consistent with the usual
global description based on the Bogoliubov transformation.
Another choice, which might be the most natural one, leads to
the local conservation of the current in an arbitrary curved background, 
provided that other interactions are absent.

Let us start with a hermitian scalar field $\phi$ in a curved background. 
The field satisfies the equation: 
\begin{equation}\label{8}
(\nabla^{\mu}\partial_{\mu}+m^2 +\xi R)\phi=0 \; ,
\end{equation}
where $R$ is the curvature.
Let $\Sigma$ be a spacelike Cauchy hypersurface. We define the scalar 
product
\begin{equation}\label{9}
(\phi_1,\phi_2)=i\cint^{\mu}\phi_1^* \apm \phi_2 \; ,
\end{equation}
where $\apm$ is the usual antisymmetric derivative. 
If $\phi_1$ and $\phi_2$ are solutions of (\ref{8}), then (\ref{9})
does not depend on $\Sigma$. It is convenient to choose coordinates 
$(t,{\bf x})$ such that $t\! =\!{\rm constant}$ on $\Sigma$. 
In these coordinates,   
the canonical commutation relations can be written as 
$[\phi(x),\phi(x')]_{\Sigma}=
[\partial_0\phi(x),\partial'_0\phi(x')]_{\Sigma}=0$ and 
\begin{eqnarray}\label{7'}
d\Sigma'^{\mu}[\phi(x),\partial'_{\mu}\phi(x')]_{\Sigma} & = & 
d^3x'\, \tilde{n}^0(x') [\phi(x),\partial'_0\phi(x')]_{\Sigma} \nonumber \\
 & = & d^3x'\, i \delta^3({\bf x}-{\bf x}') \; .
\end{eqnarray}
The label ${\Sigma}$ denotes that $x$ and $x'$ lie on ${\Sigma}$.  
The tilde on $\tilde{n}^{\mu}=|g^{(3)}|^{1/2}n^{\mu}$ denotes 
that it is not a vector, while $n^{\mu}$ is a unit vector normal 
to $\Sigma$. 

Let us first construct the particle current assuming that a 
unique (or a preferred) vacuum exists.
The field $\phi$ can be expanded as
\begin{equation}\label{10}
\phi(x)=\sum_{k}a_k f_k(x)+ a^{\dagger}_k f^*_k(x) \; , 
\end{equation}
where $(f_k,f_{k'})=-(f^*_k,f^*_{k'})=\delta_{kk'}$, 
$(f^*_k,f_{k'})=(f_k,f^*_{k'})=0$. The operators 
\begin{equation}\label{13}
a_k=(f_k,\phi) \; , \;\;\; a^{\dagger}_k=-(f^*_k,\phi)
\end{equation}
satisfy the usual algebra of lowering and raising operators. This  
allows us to introduce the vacuum as a state with the property 
$a_k|0\rangle=0$ and the operator of the total number of 
particles as
\begin{equation}\label{14}
N=\sum_{k}a^{\dagger}_k a_k \; .
\end{equation}
We also introduce the function $W(x,x')$, here defined as 
\begin{equation}\label{18r}
W(x,x')=\sum_{k} f_k(x) f^*_k(x') \; .
\end{equation}
(Later we also study different definitions of $W(x,x')$.)
For future reference, we derive some properties of this function. 
For the fields described by (\ref{10}), we find 
\begin{equation}\label{18l}
W(x,x')=\langle 0|\phi(x)\phi(x') |0\rangle \; . 
\end{equation}
From (\ref{18r}) we find $W^*(x,x')=W(x',x)$. From the fact 
that $f_k$ and $f^*_k$ satisfy Eq. (\ref{8}) we find
\begin{equation}\label{21}
(\nabla^{\mu}\partial_{\mu}+m^2 +\xi R(x))W(x,x')=0 \; ,
\end{equation}   
\begin{equation}\label{21'}
(\nabla'^{\mu}\partial'_{\mu}+m^2 +\xi R(x'))W(x,x')=0 \; .
\end{equation}   
From (\ref{18l}) and the canonical commutation relations we find 
that $f_k$ and $f^*_k$ are functions such that
\begin{eqnarray}\label{19}
& W(x,x')|_{\Sigma}=W(x',x)|_{\Sigma}, & \nonumber \\
& \partial_0\partial'_0W(x,x')|_{\Sigma}=
\partial_0\partial'_0W(x',x)|_{\Sigma}. &
\end{eqnarray}
\begin{equation}\label{20}
\tilde{n}^0\partial'_0[W(x,x')-W(x',x)]_{\Sigma}=
i\delta^3({\bf x}-{\bf x}') \; .
\end{equation}

So far nothing has been new. However, a new way of looking into the 
concept of particles emerges when (\ref{13}) is put into  
(\ref{14}) and (\ref{9}) and (\ref{18r}) are used. This leads to 
the remarkable result that (\ref{14}) can be written in the 
covariant form as 
\begin{equation}\label{15}
N=\cint^{\mu} j_{\mu}(x) \; ,
\end{equation}
where the vector $j_{\mu}(x)$ is defined as 
\begin{equation}\label{16}
j_{\mu}(x)=\cint'^{\nu} \frac{1}{2} \{W(x,x')\apm \; \apn 
\phi(x)\phi(x') + {\rm h.c.} \}.
\end{equation}
Obviously, the hermitian operator 
$j_{\mu}(x)$ should be interpreted as the    
local current of particle density. If (\ref{10}) is used 
in (\ref{16}), one can show that $j_{\mu}(x)|0\rangle =0$.

Assuming that (\ref{10}) is the usual plane-wave expansion 
in Minkowski space-time and that the $(t,{\bf x})$ coordinates 
are the usual Lorentz coordinates, we find 
\begin{equation}\label{23}
\tilde{J}_{\mu}\equiv \int d^3x \, j_{\mu}(x)=\sum_{{\bf k}} 
\frac{k_{\mu}}{\omega_{{\bf k}}} \, a^{\dagger}_{{\bf k}} 
a_{{\bf k}} \; .
\end{equation}
The quantity $k^i/\omega_{{\bf k}}$ is the 3-velocity $v^i$, so 
we find the relation $\tilde{{\bf J}}|n_q\rangle ={\bf v}n_q
|n_q\rangle$, where $|n_q\rangle$ is the state with $n_q$ 
particles with the momentum $q$. The relations of this 
paragraph support the 
interpretation of $j_{\mu}$ as the particle current.    

Note that although $j_{\mu}(x)$ is a local operator, some non-local 
features of the particle concept still remain, because (\ref{16}) 
involves an integration over $\Sigma$ on which $x$ lies. Since 
$\phi(x')$ satisfies (\ref{8}) and $W(x,x')$ satisfies 
(\ref{21'}), this integral does not depend on $\Sigma$. However, 
it does depend on the choice of $W(x,x')$. Note also that
the separation between $x$ and $x'$ in (\ref{16}) is spacelike,
which softens the non-local features because $W(x,x')$
decreases rapidly with spacelike separation. As can be explicitly
seen with the usual plane-wave modes in Minkowski space-time,
$W(x,x')$ is negligible when the spacelike separation is much larger
than the Compton wavelength $m^{-1}$.

Using (\ref{8}) and (\ref{21}), we find that 
the current (\ref{16}) possesses another remarkable property:
\begin{equation}\label{24}
\nabla^{\mu}j_{\mu}(x)=0 \; .
\end{equation}
This covariant conservation law means 
that background gravitational field does not produce particles, 
provided that a unique (or a preferred) vacuum exists.  

To explore the analogy with the current of charge, we generalize 
the analysis to complex fields, with the expansion 
\begin{eqnarray}\label{25}
\phi(x)=\sum_{k}a_k f_k(x)+ b^{\dagger}_k f^*_k(x) \; , 
\nonumber \\ 
\phi^{\dagger}(x)=\sum_{k}a^{\dagger}_k f^*_k(x)+ b_k f_k(x) \; . 
\end{eqnarray}
We introduce two global quantities:
\begin{equation}\label{26}
N^{(\pm)}=\sum_{k}a^{\dagger}_k a_k \pm b^{\dagger}_k b_k \; .
\end{equation}
Here $N^{(+)}$ is the total number of particles, while $N^{(-)}$ is 
the total charge. In a similar way we find the covariant expression 
\begin{equation}\label{27}  
N^{(\pm)}=\cint^{\mu} j^{(\pm)}_{\mu}(x) \; ,
\end{equation}
where  
\begin{eqnarray}\label{28}
j^{(\pm)}_{\mu}(x) & = & \cint'^{\nu} \frac{1}{2} \{W(x,x')\apm \; \apn
\nonumber \\
 & & 
[\phi^{\dagger}(x)\phi(x') \pm \phi(x)\phi^{\dagger}(x')] +
{\rm h.c.} \} .
\end{eqnarray}
The operator $j^{(+)}_{\mu}(x)$ is the generalization of (\ref{16}) with 
similar properties, including the non-local features. On the other hand, 
the apparent non-local features of $j^{(-)}_{\mu}(x)$ really do not  
exist, because, by using the canonical commutation 
relations, (\ref{19}) and (\ref{20}), 
$j^{(-)}_{\mu}(x)$ can be written as 
\begin{eqnarray}\label{30}
j^{(-)}_{\mu}(x) & = & i\phi^{\dagger}(x)\apm \phi(x) \nonumber \\ 
 & & + \cint'^{\nu} W(x,x')\apm \; \apn W(x',x) \; ,
\end{eqnarray}
so all non-local features are contained in the second term that 
does not depend on $\phi$. Using (\ref{25}), 
(\ref{18r}) and the orthonormality relations among the modes $f_k$, 
one can show that 
the quantity $-\langle 0|i\phi^{\dagger}(x)\!\apm\! \phi(x)| 0\rangle$ 
is equal to the second term, which reveals that the second term 
represents the subtraction of the infinite vacuum value of 
the operator represented by the first term. In other words, 
$j^{(-)}_{\mu}$ is the usual normal ordered operator of the 
charge current. 

In a way similar to the case of hermitian field, 
we find the properties $j^{(+)}_{\mu}| 0\rangle =0$ 
and $\nabla^{\mu}j^{(\pm)}_{\mu}=0$, while the generalization of 
(\ref{23}) is
\begin{equation}\label{34}
\tilde{J}^{(\pm)}_{\mu}\equiv \int d^3x \, j^{(\pm)}_{\mu}(x)=\sum_{{\bf k}}
\frac{k_{\mu}}{\omega_{{\bf k}}} (a^{\dagger}_{{\bf k}} 
a_{{\bf k}} \pm b^{\dagger}_{{\bf k}} b_{{\bf k}})\; .
\end{equation}
 
Let us now study the case in which a non-gravitational interaction 
is also present. In this case, the equation of motion is  
\begin{equation}\label{35}
(\nabla^{\mu}\partial_{\mu}+m^2 +\xi R)\phi=J \, ,
\end{equation}
where $J(x)$ is a local operator containing $\phi$ 
and/or other dynamical quantum fields. Since it describes the interaction, 
it does not contain terms linear in quantum fields. 
(For example, $J(x)$ may be the 
self-interaction operator $-\lambda\phi^3(x)$). 
We propose that even in this general case 
the currents of particles and of charge are given by the expressions 
(\ref{16}) and (\ref{28}), where $W(x,x')$ is the same function 
as before, satisfying the ``free" equations (\ref{21}) and (\ref{21'}). 
As we show below, such an ansatz leads to particle production 
consistent with the conventional approach to
particle production caused by a non-gravitational interaction. 

Note also that our ansatz for $W$ makes (\ref{30}) correct even in the 
case $J\neq 0$, because the interaction 
does not modify the canonical commutation relations. 
This implies that the charge is always 
conserved, provided that the Lagrangian possesses a global 
U(1) symmetry. On the other hand, using (\ref{21}) and (\ref{35}), we 
find
\begin{equation}\label{36}
\nabla^{\mu}j_{\mu}(x) = \cint'^{\nu} \frac{1}{2}\{ 
W(x,x')\apn J(x)\phi(x') +{\rm h.c.}\} \; ,
\end{equation}
and similarly for $j^{(+)}_{\mu}$. Note that only $x$ (not $x'$) appears 
as the argument of $J$ on 
the right-hand side of (\ref{36}), implying that $J$ plays a strictly 
local role in particle production.  

Let us now show that our covariant description (\ref{36}) of particle 
production is consistent with the conventional approach to
particle production caused by a non-gravitational interaction.   
Let $\Sigma(t)$ denote some foliation of space-time into 
Cauchy spacelike hypersurfaces. The total mean number of particles  
at the time $t$ in a state $|\psi\rangle$ is   
\begin{equation}\label{37}
{\cal N}(t)=\langle \psi |N(t)|\psi\rangle \; ,
\end{equation}
where
\begin{equation}\label{38}
N(t)=\int_{\Sigma(t)} d\Sigma^{\mu} j_{\mu}(t,{\bf x}) \; .
\end{equation}
Equation (\ref{37}) is written in the Heisenberg picture. However, 
matrix elements do not depend on picture. We introduce the 
interaction picture, where the interaction Hamiltonian is the 
part of the Hamiltonian that generates the right-hand side of 
(\ref{35}). The state $|\psi\rangle$ becomes a time-dependent 
state $|\psi(t)\rangle$, the time evolution of which is determined 
by the interaction Hamiltonian. 
In this picture the field $\phi$ satisfies the free 
equation (\ref{8}), so the expansion (\ref{10}) can be used. Since 
we have proposed that  
$W$ satisfies the free equations (\ref{21}) and (\ref{21'}), 
this implies that $N(t)$ becomes 
$\sum_{k}a^{\dagger}_k a_k$ in the interaction picture. 
Therefore, (\ref{37}) can be written as 
\begin{equation}\label{39}
{\cal N}(t)=\sum_{k}\langle \psi(t)|a^{\dagger}_k a_k|\psi(t)\rangle \; ,
\end{equation}
which is the usual formula that describes particle production caused by 
a quantum non-gravitational interaction. 

Let us now show that (\ref{16}) can also describe particle production 
by a classical time-dependent curved background, provided that we 
take a different choice for the function $W(x,x')$. When the particle 
production is described by a Bogoliubov transformation, then 
the preferred modes $f_k$ in (\ref{10}) do not exist. Instead, 
one introduces a new set of functions $u_l(x)$ for
each time $t$, such that $u_l(x)$ are positive-frequency modes 
at that time.
This means that the modes $u_l$
possess an extra time dependence, i.e. they become
functions of the form $u_l(x;t)$. These functions do not
satisfy (\ref{8}). However, the functions $u_l(x;\tau)$
satisfy (\ref{8}), provided that $\tau$ is kept fixed
when the derivative $\partial_{\mu}$ acts on $u_l$. 
To describe the local particle production, we take
\begin{equation}\label{Wcreate}
W(x,x')=\sum_l u_l(x;t)u^*_l(x';t') \; ,
\end{equation}
instead of (\ref{18r}). Since $u_l(x;t)$ do not satisfy (\ref{8}), 
the function (\ref{Wcreate}) does not satisfy (\ref{21}). 
Instead, we have
\begin{equation}\label{m2}
(\nabla^{\mu}\partial_{\mu}+m^2 +\xi R(x))W(x,x')
\equiv -K(x,x') \neq 0 \; .
\end{equation}
Using (\ref{8}) and (\ref{m2}) in (\ref{16}), we find a relation similar
to (\ref{36}):
\begin{equation}\label{m3}
\nabla^{\mu}j_{\mu}(x)=\cint'^{\nu} \frac{1}{2}\{
K(x,x')\apn \phi(x)\phi(x')+{\rm h.c.} \} \; .
\end{equation}
This local description of the particle production is
consistent with the usual
global description based on the Bogoliubov transformation. 
This is because (\ref{38}) and (\ref{16}) with (\ref{Wcreate}) 
and (\ref{10}) lead to
\begin{equation}\label{N(t)}
N(t)=\sum_l A^{\dagger}_l(t) A_l(t) \; ,
\end{equation}
where
\begin{equation}\label{e10}
A_l(t)=\sum_{k} \alpha^*_{lk}(t) a_k - \beta^*_{lk}(t) a^{\dagger}_k \; ,
\end{equation}
\begin{equation}\label{e7}
\alpha_{lk}(t)=(f_k,u_l) \; , \;\;\;\; \beta_{lk}(t)=-(f^*_k,u_l) \; .
\end{equation}
The time dependence of the Bogoliubov coefficients $\alpha_{lk}(t)$ and 
$\beta_{lk}(t)$ is related to the extra time dependence of 
the modes $u_l(x;t)$. If we assume that the
change of the average number of particles is slow, i.e. that
$\partial_t A_l(t)\approx 0$, 
$\partial_t u_l(\tau,{\bf x};t)|_{\tau=t}\approx 0$, then the Bogoliubov
coefficients (\ref{e7}) are equal to the usual Bogoliubov 
coefficients. This approximation is nothing else but the
adiabatic approximation,
which is a usual part of the convential
description of particle production \cite{park}. 
 
In general, 
there is no universal natural choice for the modes $u_l(x;t)$. 
In particular, in a given space-time, the choice of the natural 
modes $u_l(x;t)$ may depend on the observer. If different 
observers (that use different coordinates) use different modes for the 
choice of (\ref{Wcreate}), then 
the coordinate transformation alone does not describe how  
the particle current is seen by different observers. In this sense, 
the particle current is not really covariant. There are also 
other problems related to the case in which different 
observers use different modes \cite{nikolmpl}. 

Since the covariance was our original aim, it is desirable to 
find a universal natural choice of $W(x,x')$, such that,
in Minkowski space-time, it 
reduces to the usual plane-wave expansion (\ref{18r}). 
Such a choice exists. This is the function $G^+(x,x')$ 
satisfying (\ref{21}) and (\ref{21'}), calculated in a 
well-known way from the Feynman propagator $G_F(x,x')$ \cite{bd,dewitt}. 
The Feynman propagator $G_F(x,x')$, 
calculated using the Schwinger-DeWitt method, is unique, provided that a 
geodesic connecting $x$ and $x'$ is chosen.  
When $x$ and $x'$ are sufficiently close to one another, 
then there is only one such geodesic. In this      
case, the adiabatic expansion \cite{bd,dewitt} of the 
Feynman propagator can be used.
For practical calculations, the adiabatic expansion may be sufficient because 
$G^+(x,x')$ decreases rapidly with spacelike separation, 
so the contributions from large spacelike separations may be 
negligible.

To define the exact unique particle current, we need a natural 
generalization of $G_F(x,x')$ to the case with more than one 
geodesic connecting $x$ and $x'$. The most natural choice is the  
two-point function $\tilde{G}_F(x,x')$ defined as the average over 
all geodesics connecting $x$ and $x'$. Assuming that there are $N$ 
such geodesics, this average is 
\begin{equation}\label{average}
\tilde{G}_F(x,x')=N^{-1} \sum_{a=1}^{N} G_F(x,x';\sigma_a) \; ,
\end{equation}
where $G_F(x,x';\sigma_a)$ is the Feynman propagator $G_F(x,x')$ calculated 
with respect to the geodesic $\sigma_a$. Eq. (\ref{average}) can 
be generalized even to the case with a continuous set of geodesics 
connecting $x$ and $x'$. This involves a technical subtlety related 
to the definition of measure on the set of all geodesics, 
which we shall explain in detail elsewhere.

Since the function $\tilde{G}^+(x,x')$ calculated from 
$\tilde{G}_F(x,x')$ always satisfies (\ref{21}) and (\ref{21'}), it 
follows that the corresponding particle current is conserved and 
does not depend on the choice of $\Sigma$, provided that the field 
satisfies (\ref{8}). Note that this definition of
particles does not
always correspond to the quantities detected
by ``particle detectors" of the Unruh-DeWitt type \cite{unruh,dewitt2}.
Instead, this number of particles is determined by a well-defined
hermitian operator that does not require a model of a particle
detector, just as is the case for all other observables
in quantum mechanics. Moreover, since (\ref{16}) does not require
a choice of representation of field algebra, the definition
of particles based on $\tilde{G}_F$ does not require the choice of
representation either. This allows us to treat the particles in
the framework of algebraic quantum field theory in curved
background \cite{wald}.
Furthermore, as noted by Unruh
\cite{unruh}, only one definition of particles can correspond to the
real world, in the sense that their stress-energy contributes to
the gravitational field. Since the definition of particles based on 
$\tilde{G}_F$ is universal, unique and really covariant, it might 
be that these are the particles that correspond to the real world.
Besides, as we shall discuss elsewhere,
it seems that these particles might correspond
to the objects detected by real detectors
(such as a Wilson chamber or
a Geiger-M\"{u}ller counter) in real experiments.
  
In this letter we have constructed the operator describing the 
local current of particle density, which allows us to treat 
the concept of particles in quantum field theory in a local and 
generally covariant manner. This operator is not unique, but 
depends on the choice of the two-point function $W(x,x')$. 
Different choices correspond to different definitions of particles. 
In particular, various choices based on (\ref{Wcreate}) correspond 
to particle production by the gravitational field. 
This local description of particle production is consistent with 
the conventional global description based on the Bogoliubov 
transformation. Similarly, various choices based on (\ref{18r}) 
give a local description of particle content in various inequivalent 
representations of field algebra. 
 
We have seen that a
particularly interesting choice of $W(x,x')$ is that based 
on the Feynman propagator $\tilde{G}_F(x,x')$, because this 
choice might correspond to the most natural 
universal definition of particles for 
quantum field theory in curved background.  
It is tempting to interpret these particles as
real particles.
If this interpretation is correct, then 
classical gravitational backgrounds do not produce real particles. 
There are also other indications that classical gravitational 
backgrounds might not produce particles 
\cite{padmprl,belin,nikol123}. 
Even if 
the particles based on $\tilde{G}_F$ 
do not correspond to real physical particles in general, 
it is interesting to ask about the physical meaning of this 
time-independent hermitian observable that corresponds to physical 
particles at least in Minkowski space-time.
 
Note finally that        
our definition of the particle current can be generalized to the case 
of scalar and spinor fields in classical electromagnetic backgrounds.
Again, a two-point function can be chosen such that the particle 
production is described in a way consistent with the Bogoliubov-transformation 
method, but the requirement of gauge invariance leads to a conserved 
current, 
indicating that classical electromagnetic backgrounds also 
might not produce particles. Only the quantized electromagnetic field 
(having a role similar to $J(x)$ in (\ref{35})) may cause particle 
production in this formalism, which is in agreement with some other 
results \cite{nikol123}.
This will be discussed in more detail elsewhere.
 
This work was supported by the Ministry of Science and Technology of the
Republic of Croatia under Contract No. 00980102.

\end{document}